\documentclass{article}
\usepackage{spconf,amsmath,graphicx}
\usepackage{color}

\usepackage{amsmath,amssymb,amsfonts}
\usepackage{subfig}
\usepackage{graphicx}
\usepackage{textcomp}
\usepackage{graphicx}
\usepackage{float}
\usepackage{url}
\usepackage{longtable}
\usepackage{booktabs}
\usepackage{threeparttable}

\usepackage{multirow}

\usepackage{algorithmicx}
\usepackage[linesnumbered,ruled,vlined]{algorithm2e}
\usepackage{algpseudocode}
\usepackage{amsmath}
\usepackage{enumitem}
\setlist{nosep, leftmargin=14pt}

\usepackage{mwe} 

\title{Anatomy-guided fiber trajectory distribution estimation for cranial nerves tractography}
\name{\normalsize Lei Xie$^{1}$, Qingrun Zeng$^{1}$, Huajun Zhou$^{2}$, Guoqiang Xie$^{3}$, Mingchu Li$^{4}$, Jiahao Huang$^{1}$, Jianan Cui$^{1}$, Hao Chen$^{2}$, Yuanjing Feng$^{1,*}$\thanks{ $^{*}$Corresponding authors. e-mail:fyjing@zjut.edu.cn}}
\address{$^{1}$ ************************************************ \\
   $^{2}$ ************************************************\\
$^{3}$ ************************************************}
\address{\normalsize$^{1}$ College of Information Engineering, Zhejiang University of Technology \\
\normalsize	$^{2}$ Department of Computer Science and Engineering, The Hong Kong University of Science and Technology\\
	\normalsize$^{3}$ Department of Neurosurgery, Nuclear Industry 215 Hospital of Shaanxi Province\\
\normalsize$^{4}$Department of Neurosurgery, Capital Medical University Xuanwu Hospital}
\begin{document}
%
\maketitle
\begin{abstract}
Diffusion MRI tractography is an important tool for identifying and analyzing the intracranial course of cranial nerves (CNs). 
However, the complex environment of the skull base leads to ambiguous spatial correspondence between diffusion directions and fiber geometry, and existing diffusion tractography methods of CNs identification are prone to producing erroneous trajectories and missing true positive connections. To overcome the above challenge, we propose a novel CNs identification framework with anatomy-guided fiber trajectory distribution, which incorporates anatomical shape prior knowledge during the process of CNs tracing to build diffusion tensor vector fields. We introduce higher-order streamline differential equations for continuous flow field representations to directly characterize the fiber trajectory distribution of CNs from the tract-based level.
The experimental results on the vivo HCP dataset and the clinical MDM dataset demonstrate that the proposed method reduces false-positive fiber production compared to competing methods and produces reconstructed CNs (i.e. CN II, CN III, CN V, and CN VII/VIII) that are judged to better correspond to the known anatomy.
\end{abstract}
\section{Introduction}
\label{sec:intro}
The cranial nerves (CNs), twelve pairs of nerves emerging directly from the brain, hold paramount importance in regulating various sensory and motor functions throughout the body~\cite{zolal2016comparison}. Accurately identifying CNs is of immense significance as it allows for precise diagnosis and targeted treatment of neurological disorders. However, complete and accurate CNs reconstruction is challenging due to the skull base environment and complex fiber geometry that can lead to abnormal diffusion signals~\cite{xie2023cntseg}. 
\begin{figure}[]
	\centering
	\includegraphics[width=0.39\textwidth]{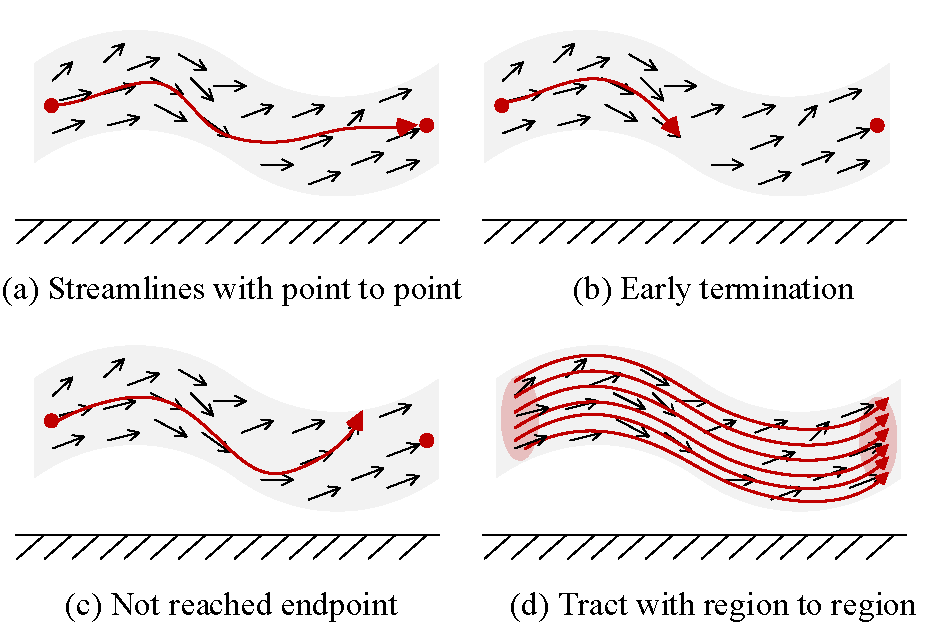}
	\caption{Schematic representation of CNs fiber streamlines.}
	\label{fig:pic1}
\end{figure}

Diffusion MRI (dMRI) tractography has been applied successfully for the identification of the CNs with the advantage of non-invasive in vivo reconstruction of the three-dimensional trajectory of CNs~\cite{jacquesson2019overcoming}\cite{yoshino2016visualization}. The researchers utilized suitable fiber tractography methods for CNs identification to filter CNs from streamline results using the regions of interest (ROIs) selection strategy, where ROIs are performed by manually labeling specific anatomical regions in MRI and dMRI by trained experts~\cite{he2021comparison}. Deterministic methods (DT)~\cite{basser2000vivo}, probabilistic methods~\cite{behrens2007probabilistic}, and unscented kalman filter (UKF) algorithms~\cite{malcolm2010filtered} will generally be used to identify CNs tract, and these methods compute the current fiber orientation distribution at each new location, indicating the direction of propagation of the tracking method's next step through a specific strategy. However, the complex fiber geometry of the CNs and signal interference increase the error when performing point-to-point streamline generation (Fig.~\ref{fig:pic1}(a)) using fiber orientation distribution function peaks, which can lead to early interruptions (Fig.~\ref{fig:pic1}(b)) and failure to reach the end point(Fig.~\ref{fig:pic1}(c)) during the tracking process of the above algorithms. Thereby, this motivates us to propose the research question of this paper: \textit{How to directly identify CNs by incorporating the overall trajectory distribution at the tract-based level to reduce false positive streamlines generation, i.e., track with region to region as illustrated in Fig.~\ref{fig:pic1}(d)?}

To this end, we develop a novel CNs tractography framework with fiber trajectory distribution (FTD) using anatomical shape priors. To describe the CNs tract from region to region, we define an anatomy-guided fiber trajectory distribution function based on a higher-order streamline differential equation in the measured diffusion tensor vector field. We build the anatomical shape priors by estimating CNs orientation from the obtained CNs centerlines, and this prior will be used as an anatomical constraint on the optimization model to estimate the fiber trajectory distribution function.
The experimental results demonstrate that the proposed method can successfully identify five pairs of CNs (i.e. CN II, CN III, CN V, and CN VII/VIII) compared to state-of-the-art methods and shows better spatial congruence with fiber geometry.
\section{Background on the FTD function}
Previously, we defined an asymmetric FTD function on the cube of each voxel, which is the fiber trajectory that passes through the voxel~\cite{feng2020asymmetric}. We denote a white matter fiber bundle trajectory as a curve flow described by a vectorial field. The definition of an FTD function is as follows:
 
Consider $\small {v}(x,y,z) = {[{v_x}(x,y,z),{v_y}(x,y,z),{v_z}(x,y,z)]^T}$ the field vector of diffusion tensor at position $(x,y,z)$. The fiber flow can then be parameterized by a set of streamlines $\mathcal{S} = \left\{s_i, i = 1,...,n\right\}$ that satisfied the following properties: 
\begin{enumerate}
	\item [i)] The tangent vector at each point $(x,y,z)$ of fiber path $s_i$ equals to the field vector $v(x,y,z)$, that is,
	\begin{equation}\label{eq:v}
	\frac{{\partial {s_i}}}{{\partial x}} = {v_x}(x,y,z),\frac{{\partial {s_i}}}{{\partial y}} = {v_y}(x,y,z),\frac{{\partial {s_i}}}{{\partial z}} = {v_z}(x,y,z).
	\end{equation}
	\item [ii)]The streamlines satisfied the fiber trajectories don't intersect, which is defined by fliud differential equation,
	\begin{equation}
	\frac{{dx}}{{v_x}(x,y,z)}{\rm{ }} = {\rm{ }}\frac{{dy}}{{v_y}(x,y,z)} = \frac{{dz}}{{v_z}(x,y,z)}.
	\label{eq6}
	\end{equation}

\item [iii)]According to the theory of continuous incompressible fluids, the spatial continuity of fiber flow $\Omega$ trajectories is described by introducing the concept of divergence,
\begin{equation}\label{eq:divs}
	div\Omega = \frac{{\partial {{v_x}}}}{{\partial x}} + \frac{{\partial {{v_y}}}}{{\partial y}} + \frac{{\partial {{v_z}}}}{{\partial z}},
\end{equation}
when the fibers terminate in the terminate area of field, the divergence satisfies $div\Omega = 0$.

\end{enumerate}
\section{METHODOLOGY}
The FTD function is defined on neighboring voxels based on the streamline differential equation with the tangent vector approximated by ternary quadratic polynomials. Instead, we introduce the higher-order streamline differential equation to directly reconstruct more complex CNs trajectories from the tract-based level using anatomical priors as constraints.
\subsection{Definition of tract-based FTD function}
In order to define the tract-based FTD function directly, we introduce the higher-order streamline differential equation with the tangent vector approximated by the $N^{th}$-order polynomial, which is expressed as
\begin{equation}\label{eq:order}
	\mathcal{F}(x,y,z) = \sum\limits_{i = 0}^N {\sum\limits_{j = 0}^{N - i} {\sum\limits_{k = 0}^{N - i - j} {{a_{i,j,k}}{x^i}} } } {y^j}{z^k},
\end{equation}
where $a_{ijk}$ are the coefficients of the polynomial. The diffusion vector ${v}(x,y,z)$ at each position $(x,y,z)$ can be expressed as,
\begin{equation}\label{eq:VF}
\small
v(x,y,z) = {\left[ {{{\cal F}_x}(x,y,z),{{\cal F}_y}(x,y,z),{{\cal F}_z}(x,y,z)} \right]^T}.
\end{equation}
Combining with Eq.~\ref{eq:order} and Eq.~\ref{eq:VF}, the diffusion vector can be transformed into,
\begin{equation}\label{eq:ac}
		{v}(x,y,z) = \mathcal{A} \cdot \mathcal{D}(x,y,z)^T,
\end{equation}
where $\mathcal{D}(x,y,z) \in {\mathbb{R}^{1 \times M}}$ and $\mathcal{A} \in {\mathbb{R}^{3 \times M}}$ denote as,
\begin{equation}\label{eq:c}
	\mathcal{D}(x,y,z) = {\left\{{x^i}{y^j}{z^k}\right\}_{i \in [0,N],j \in [0,N - i],k \in [0,N - i - j]}},
\end{equation}
\begin{equation}\label{eq:a}
	\mathcal{A} = {\left\{ {a_{i,j,k}^x,a_{i,j,k}^y,a_{i,j,k}^z} \right\}_{i \in [0,N],j \in [0,N - i],k \in [0,N - i - j]}},
\end{equation}
where $a_{i,j,k}^x,a_{i,j,k}^y,a_{i,j,k}^z$ represent the coefficients of the polynomial in the $x,y,z$ axis, and \begin{footnotesize}$M=(N+1)(N+2)(N+3)/6$\end{footnotesize}. The estimation of ternary form aims to find the optimal global trajectory by fitting the estimated directions from fiber orientation distribution (FOD) with the divergence restriction.

\subsection{Tract-based FTD function estimation with anatomically constrained optimization model}
With the definition of tract-based FTD function, finding the optimal CNs tract $S$ can be simplified as the estimation of coefficient matrix $\mathcal{A}$, which can be estimated using diffusion images by minimizing the following cost function:
\begin{equation}
	\begin{split}
		\mathop {\min }\limits_\mathcal{S} \iiint_{{\mathrm{CN}_{\Omega}}}{\left\| {\Phi ({v}(x,y,z)) - {v}(x,y,z)} \right\|_2^2dxdydz
		}  \\ s.t. \frac{{\partial {{v_x}}}}{{\partial x}} + \frac{{\partial {{v_y}}}}{{\partial y}} + \frac{{\partial {{v_z}}}}{{\partial z}}  = 0.
	\end{split}
	\label{eq14}
\end{equation}
where $\Phi ({v}(x,y,z))$ is the probability that fiber trajectory is actually passes through the FOD at point $(x,y,z)$, and ${\mathrm{CN}_{\Omega}} \in {\mathbb{R}^3}$ is the bundle pathway from region to region. 

To achieve overall optimisation of cranial nerves, we add constraints by introducing an anatomical prior. First, we formulate centerline extraction problem inside an object ${\mathrm{CN}_{\Omega}}$ as finding a path $\mathcal{C}(\tilde s)$ ($\tilde s$ being arc length) between two points $p_1$ and $p_2$ for which the following energy functional
\begin{equation}\label{eq:centerline}
	{\mathrm{E}_{centerline}}(\mathcal{C}) = \int_{0 = {\mathcal{C}^{ - 1}}({p_1})}^{L = {\mathcal{C}^{ - 1}}({p_2})} {\mathcal{H}(\mathcal{C}(\tilde s))d\tilde s}
\end{equation}
is minimal, where $\mathcal{H}(x)$ is a scalar field which is lower for more internal points~\cite{piccinelli2009framework}, for example a decreasing function of the distance transform associated with ${\mathrm{CN}_{\Omega}}$, defined as 
\begin{equation}
	\mathrm{DT}(x) = \mathop {\min }\limits_{ y \in \partial \mathrm{CN}_{\Omega}} {\  || x - y \  ||}
\end{equation}
where ${\  || \cdot  ||}$ denotes the Euclidean distance, and $\partial \mathrm{CN}_{\Omega}$ the boundary of object ${\mathrm{CN}_{\Omega}}$. 
By choosing $\mathcal{H}(x) = {\mathrm{DT}^{-1}}(x)$, centerlines defined
as in Eq.~\ref{eq:centerline} lie on the medial axis $\mathrm{MA}(\mathrm{CN}_{\Omega})$, which can be identified with the ridges of its distance transform 
\begin{equation}\label{eq:MA}
	\begin{split}
		\mathrm{MA}(\mathrm{CN}_{\Omega})&= \{x\in \mathrm{CN}_{\Omega} \  | \ \exists\mathrm{dir} \ \mathbf{n}, \ \exists\bar\varepsilon>0: \\
		& \ \forall \varepsilon\in(0,\bar\varepsilon),\mathrm{DT}(x)\geq\mathrm{DT}(x+\varepsilon\mathbf{n} \}.
	\end{split}
\end{equation}
Once the $\mathrm{MA}(\mathrm{CN}_{\Omega})$ is obtained by using toolkit VMTK~\cite{piccinelli2009framework}, we calculate the angle $\theta$ between the FOD and the normal vector $\mathbf{n}$ the cross-section of each voxel on the $\mathrm{MA}(\mathrm{CN}_{\Omega})$ where the current point $(x,y,z)$ is located.
Each selected direction from the fODF peaks had the minimum angle $\theta$ with the normal vector $\mathbf{n}$ at the current voxel. Finally, we traverse the points on the entire centerline to obtain the orientation estimation $\mathcal{V}(x,y,z)$ of the CNs. 
To achieve spatial continuity of the fiber, we take the above anatomical prior as a constraint and the optimization model in Eq.~\ref{eq14} is simplified as 
\begin{equation}\label{eq:min}
\begin{split}
\min\limits_\mathcal{A} \sum\limits_{}^{} {_\Omega \left\| {\mathcal{V}(x,y,z)  - v(x,y,z)} \right\|} _2^2, s.t. div\Omega = 0.
\end{split}
\end{equation}
The coefficient matrix $\mathcal{A}$ from Eq.~\ref{eq:min} can be calculated through an interior-point algorithm using fmincon of the MATLAB function.
\subsection{CNs tractography with tract-based FTD function}
\begin{algorithm}[b]
	\caption{The CNs tractography algorithm.}
	\label{Algorithm:1}
	\KwIn{fODF peak, mask ${\mathrm{CN}_{\Omega}}$, seeds $\beta _i$, step $\lambda$.
	}
	\KwOut{CNs tract  $S = \left\{ {{s_i},i = 1, \cdots ,n } \right\}$.} 
	\For{$i = 1 \to n$}{
		set as starting point of $s_i$\\
		$\eta (0) = {\beta _i}$, $t=0$\\
		\While{$\eta (t) \in {\mathrm{CN}_{\Omega}}$}{
			$v(\eta (t)) = \mathcal{A} \cdot \mathcal{D}(\eta (t))$\\
			$v_g(\eta (t)) \leftarrow $sampled from the $GD$ function\\
			$\eta (t+1) = RungeKutta({\rm{ }}\eta (t),v_g(\eta (t)),\lambda )$\\
			$s_i = \left\{ {\eta (0), \cdots ,\eta (t+1) } \right\}$\\
			$t = t+1$\\	
		}	
	}\textbf{Return} CNs tract $S = \left\{ {{s_i},i = 1, \cdots ,n } \right\}$
\end{algorithm}
For each CNs tract, a specific CNs trajectory distribution is estimated for fiber tractography. 
Once the coefficient $\mathcal{A}$ is obtained by the least squares method, we use Eq.~\ref{eq:ac} to obtain the flow field vector $v(\eta (t))$ at the current point $\eta (t)$. In order to learn a more complete distribution of orientations for each voxel, we obtain the new resample orientations $v_g(\eta (t))$ by using a Gaussian distribution (GD) function centered on $v(\eta (t))$ with a fixed standard deviation. At each step of the tracking process, the direction is drawn from the Gaussian distribution~\cite{wasserthal2019combined}. 
Finally, we solve the higher-order streamline differential equation using the fourth-order Runge-Kutta method~\cite{basser2000vivo}. See Algorithm.~\ref{Algorithm:1} for a detailed implementation. 
\section{EXPERIMENTS and results}\label{sec:EXPERIMENTS}
\subsection{Experimental settings}
We evaluate the proposed method both on the public HCP dataset~\cite{sotiropoulos2013advances} and the public MDM clinical  dataset~\cite{tong2019reproducibility}. The experimental results are evaluated by comparing our proposed method with iFOD1, DT, UKF, PTT~\cite{aydogan2020parallel}, and FTD. FOD were estimated using constrained spherical deconvolution. The best-performing parameters for CN V are as shown in Table.~\ref{tab1}, and the other CNs are given in \cite{he2021comparison}\cite{zeng2023automated}\cite{huang2022automatic}. For all tractography methods, the individual CNs were obtained by screening with the same ROIs~\cite{xie2023cntseg}, which were manually drawn by the neuroradiological experts. In this paper, we use the spatial overlap~\cite{2020Creation} and fiber distance (HD and AHD)~\cite{zeng2023automated}\cite{xie2022semi}\cite{xie2020anatomical} to assess the CNs reconstruction performance of each tractography method with manually identified CNs. The reference data is obtained by setting the ROIs and ROAs for screening, and finally the neurosurgeon manually removes false positive fibers.
\begin{table}[!h]
	\centering
	\caption{Tractography parameters used in the experiments.}
	\label{tab1}
		\resizebox{0.42\textwidth}{!}{
	\begin{tabular}{cc}
		\hline\hline
		Methods          & Tractography Parameters\\
		\hline
		iFOD1          &\textit{cutoff} = 0.05, \textit{angle} = 12, \textit{stepsize} = 0.5\\
		DT         & \textit{cutoff} = 0.05, \textit{angle} = 40, \textit{stepsize} = 0.5  \\ 
		\multirow{2}{*}{UKF}        &\textit{Qm} = 0.001, \textit{Ql} = 50, \textit{stoppingFA} = 0.05  \\ 
		      & \textit{stoppingThreshold} = 0.06, \textit{seedingFA} = 0.06 \\ 
		      PTT       & default parameters in~\cite{aydogan2020parallel}   \\ 
		      FTD       &  \textit{stepsize} = 0.3, others are default in~\cite{feng2020asymmetric} \\ 
		      OUR       &  \textit{stepsize} = 0.3, $N=4$ \\ 
		\hline\hline
	\end{tabular}}
\end{table}
\subsection{Performance on HCP dataset}
\begin{figure}[t]
	\centering
	\includegraphics[width=0.46\textwidth]{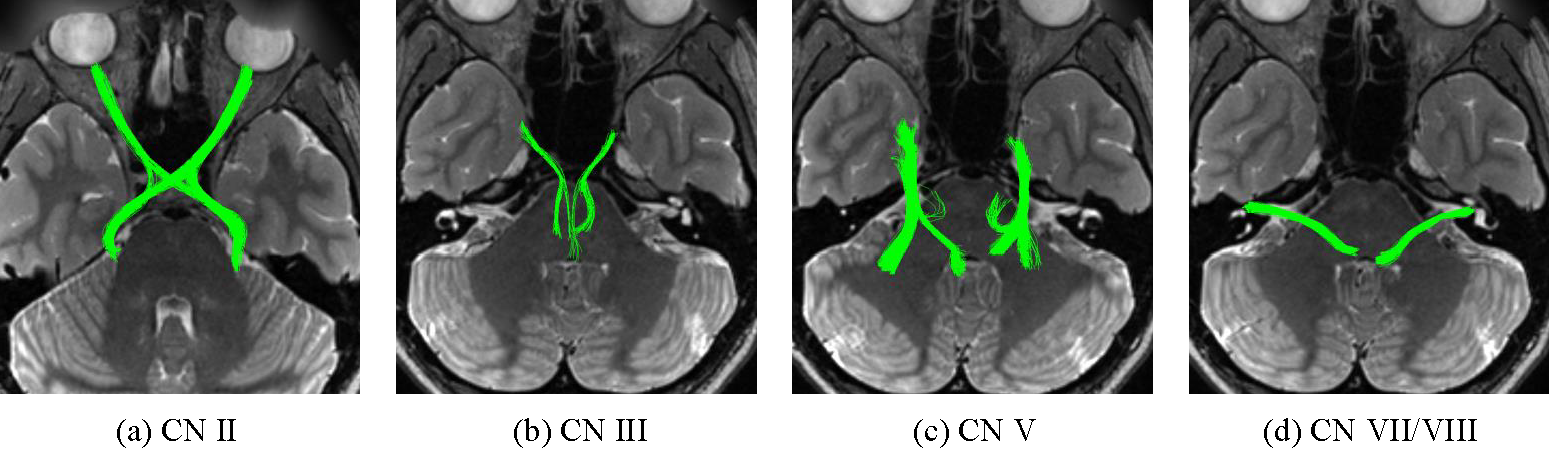}
	\caption{Visualization of the fiber trajectory of the CN II, CN III, CN V, and CN VII/VIII in HCP \#100307 subject, overlaid on T2w images.}
	\label{fig:hcpresults1}
\end{figure}

\begin{table}[b]
	\centering
	\caption{The CN V reconstruction quantitative comparison on HCP dataset.}
	\label{tab:2}
	\resizebox{0.46\textwidth}{!}{%
		\begin{tabular}{ccccccc}
			\hline\hline
			& iFOD1           & DT & UKF            & PTT          & FTD & OUR \\ \hline
			Spatial Overlap [\%] $\uparrow$ & 70.45         & 75.59        & 81.29       & 76.22       & 80.12        & \textbf{91.31}       \\ 
			Fiber Distance (\textit{HD}) $\downarrow$ & 25.76         & 30.56        & 18.83          & 24.68       & 31.57          & \textbf{16.72 }         \\ 
			Fiber Distance (\textit{AHD}) $\downarrow$ & 1.51          & 2.59        & 1.11         & 1.67      & 2.68        & \textbf{0.84}          \\ 
			\hline\hline
		\end{tabular}
	}
\end{table}
In this experiment, we test the proposed method on 10 randomly selected subjects from the HCP dataset: 18 base images with b-values = 0 s/mm$^2$ and 270 gradient directions. b = 1000, 2000, and 3000 s/mm$^2$, TR = 5520 ms, TE = 89.5 ms, matrix size = 145 × 174 × 145, resolution = 1.25 × 1.25 × 1.25 mm$^3$ voxels.
Fig.~\ref{fig:hcpresults1} gives the tractography results of the CN II, CN III, CN V, and CN VII/VIII using our proposed method. We can see that the five pairs of the CNs identified are anatomically consistent from the fiber trajectory.
\begin{figure}[t]
	\centering
	\includegraphics[width=0.46\textwidth]{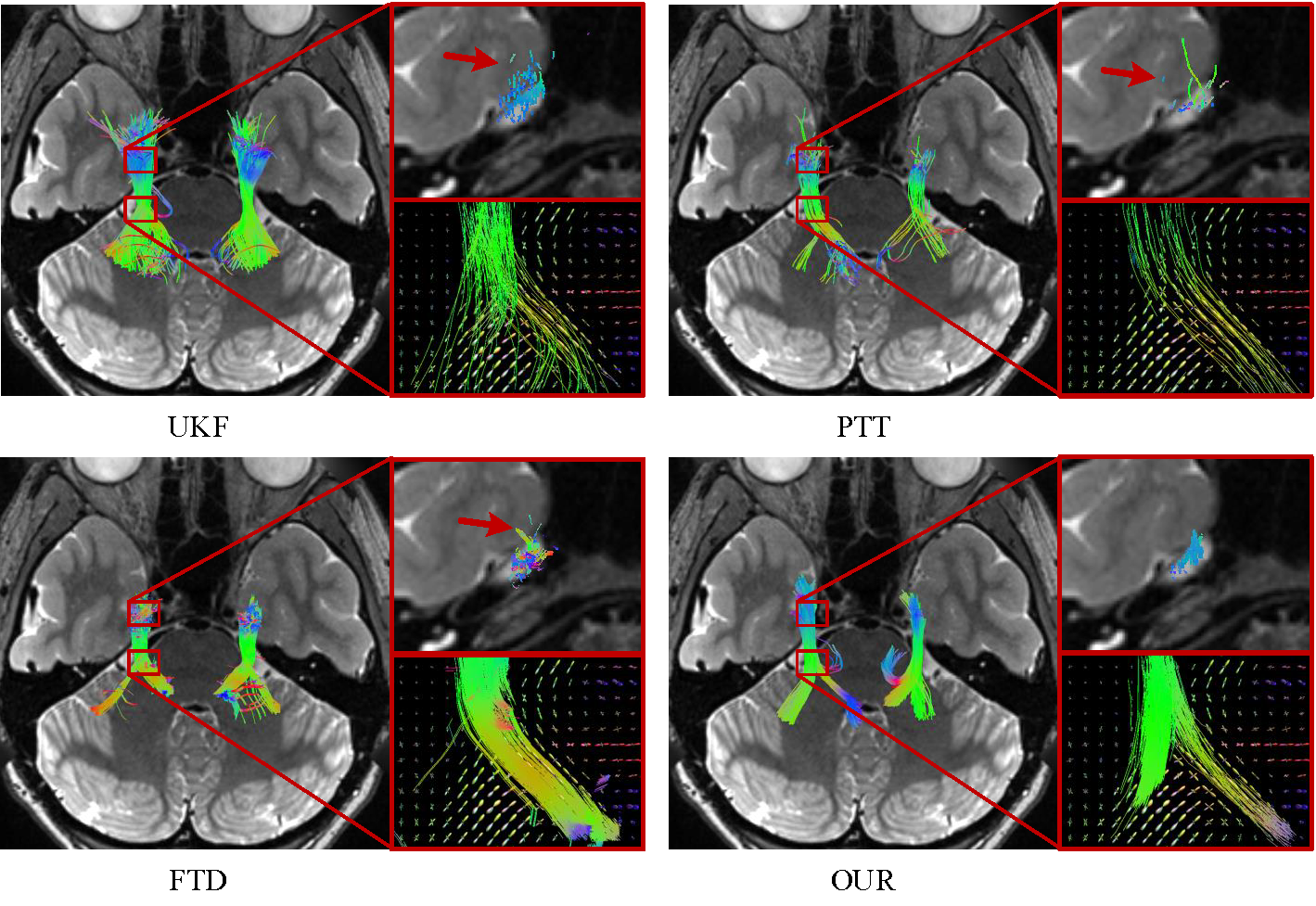}
	\caption{
		Visual comparison of the CN V fiber trajectory obtained on the HCP dataset using UKF, PTT, FTD, and the proposed method. Red arrows point to false-positive fiber generated in the temporal lobe.}
	\label{fig:hcpresults}
\end{figure}

Further, we choose the structurally complex trigeminal nerve (CN V) for a more detailed visual comparison. Table.~\ref{tab:2} gives the spatial overlap and fiber distance of the CN V reconstruction using different methods, and we can see that the validation metrics of the proposed method are all better than the other methods. In terms of fiber distance metrics, the CN V reconstructed by our method is closer to manually identified CN V.
We use subject \#100307 from the HCP dataset to visually compare the CNs reconstruction performance of our proposed method. As shown in Fig.~\ref{fig:hcpresults}, for each sub-figure, inset images are provided for better visualization of the region through which the CN V passes and the fiber orientations corresponding to the generated streamlines. We can observe that the reconstructed CN V using our method is spatially consistent with the true fiber orientation, and false positive fibers are effectively reduced. It is worth noting that from each of the inset images, we can see that our method generates streamlines that are more consistent with the true fiber orientation where the fiber diverges.
\begin{figure}[t]
	\centering
	\includegraphics[width=0.4\textwidth]{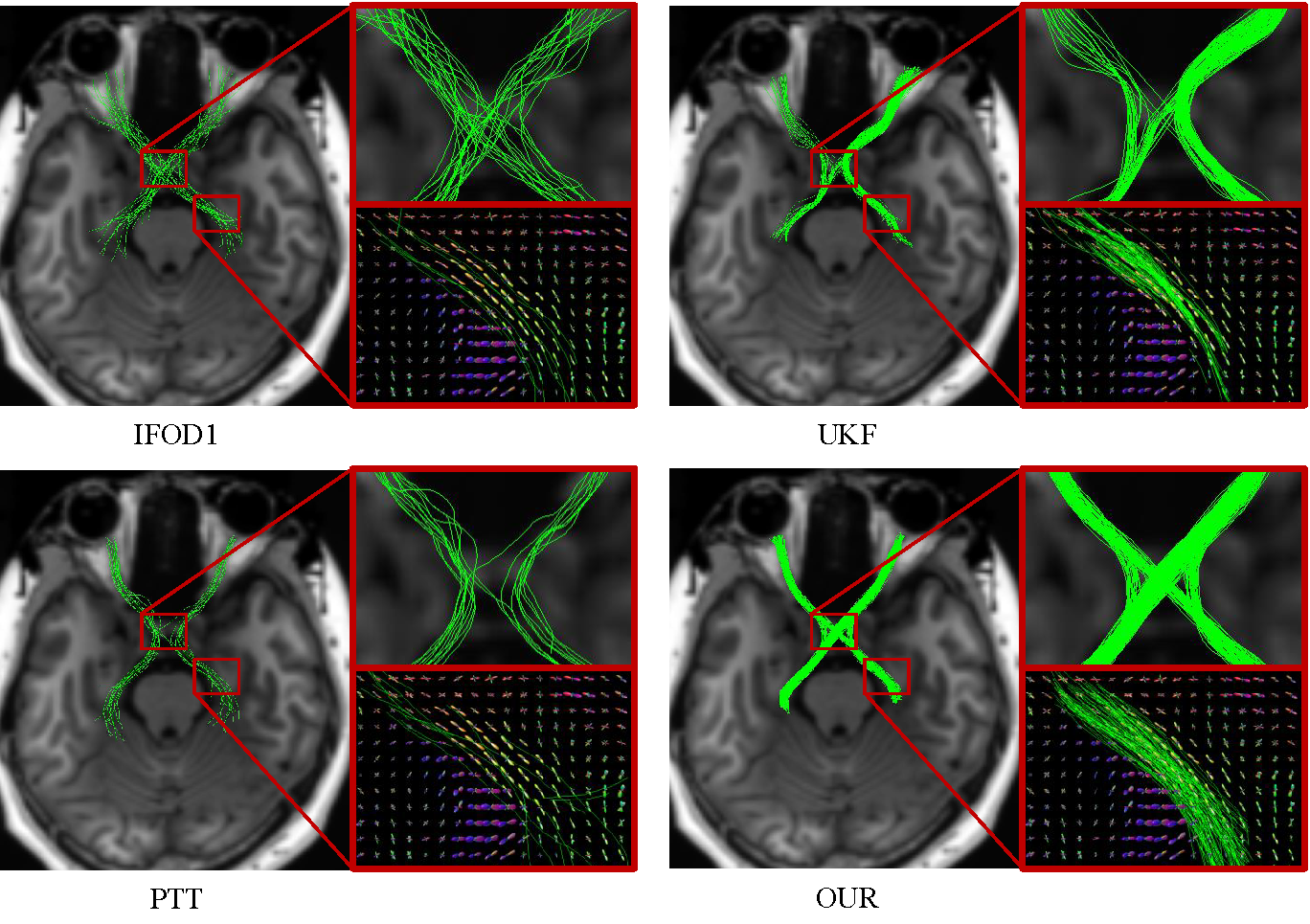}
	\caption{Visual comparison of the CN II fiber trajectory obtained on the MDM dataset using IFOD1, UKF, PTT, and the proposed method.}
	\label{fig:MDMresults}
\end{figure}
\subsection{Performance on MDM dataset}
In this experiment, we evaluate the performance of the proposed method using one subject from the MDM dataset~\cite{tong2019reproducibility}: 6 base images with b-values = 0 s/mm$^2$ and 90 gradient directions with other three b-values of 1000, 2000, and 3000 s/mm$^2$ and 1.5 mm$^3$ isotropic spatial resolution. We choose the CN II with crossed and fan-shaped fibers to validate the performance of our method on clinical MDM data. Fig.~\ref{fig:MDMresults} gives the results of the reconstruction of the CN II and the direction of FODs corresponding to the generated streamlines. We can see that the fiber streamlines generated by the proposed method completely cover the region where the real CN II passes through, and it is effective in reducing the generation of false-positive fibers compared with other methods. 
\section{Discussion and Conclusion}\label{sec:Discussion}
In this paper, we propose a novel CNs identification framework with anatomy-duided fiber trajectory distribution, which incorporates pre-existing anatomical knowledge into the process of tracing cranial nerves, enabling the construction of diffusion tensor vector fields. The experimental results on the HCP dataset and the MDM dataset demonstrate that the proposed method exhibits performance in CNs identification that is highly comparable to other existing methods. Regarding the generation of anatomical shape priors, we can use the extraction of the median axis or can utilize individualised CNs atlas to obtain orientation priors. The limitation of the proposed method is when FOD or peaks are affected by factors such as noise, artifacts, pathological conditions, or low-quality datasets, it may experience some impact. The future study will focus on modeling the optimal transport in the flow field to solve the problem of false positive fibers during the tracking process.
\section{COMPLIANCE WITH ETHICAL STANDARDS}
This study was conducted retrospectively using the public HCP dataset~\cite{sotiropoulos2013advances} and the MDM dataset~\cite{tong2019reproducibility}. No ethical approval was required.
\section{ACKNOWLEDGEMENTS}
This work was sponsored in part by the National Natural Science Foundation of China (Grant No. 62002327, 61976190, U22A2040), and Natural Science Foundation of Zhejiang Province (Grant No. LQ23F030017).
\small
\bibliographystyle{IEEEbib}
\bibliography{strings,refs}

\end{document}